# Analytic model for grain-boundary segregation energies in metal polycrystal

Hao Wu[#], Xin Li[#], Wang Gao*, and Qing Jiang*

Key Laboratory of Automobile Materials, Ministry of Education, Department of Materials Science and Engineering, Jilin University, 130022, Changchun, China

[#]These authors contributed equally

Solute segregation at grain boundaries (GBs) of polycrystals strongly impacts the mechanical properties of metals including strength, fracture, embrittlement, and corrosion[1-5]. However, the complexity of GB structures and the large chemical space of solutes and matrices impede the understanding of segregation. Herein, we identify a physical-based determinant, by unifying the effects of plastic strain and bonding breaking, for determining the segregation energies at GBs. By further combining with the usual coordination number, atomic radius of solutes and matrices, and cohesive energy of matrices, we build an analytic framework to predict segregation energies of polycrystal GBs across various solutes and matrices. These findings indicate an unusual Coulombic-like and localized nature of the bonding at polycrystal GBs and bulk metallic glasses (BMGs). Our scheme not only uncovers the coupling rule of solutes and matrices for GB segregation in polycrystals, but also provides an effective tool for the design of high-performance alloys.

The experimental measurements are difficult to quantify the segregation energy, except with the help of some classical models, such as Langmuir-McLean types of segregation isotherm[5-7]. However, the accuracy of these empirical models is limited and the generalization in polycrystalline metals remains to be proven[8]. Some numerical methods, such as density functional theory (DFT)[8-10], are commonly utilized to study GB segregation energies, but are mainly applicable to the bicrystal GBs with less insight into the intrinsic determinants of polycrystal GBs. Based on molecular dynamics (MD) calculations[11,12], machine learning (ML) methods are used to predict the segregation energies of polycrystal GBs by combining with the smooth overlap of atomic positions (SOAP) method[13,14]. However, the ML approaches are often black-box techniques with unclear physical meanings and inexplicit mathematical functions.

To construct the overall physical picture of solute segregation at polycrystal GBs, we study the segregation energies of over 200 polycrystal structures incorporating

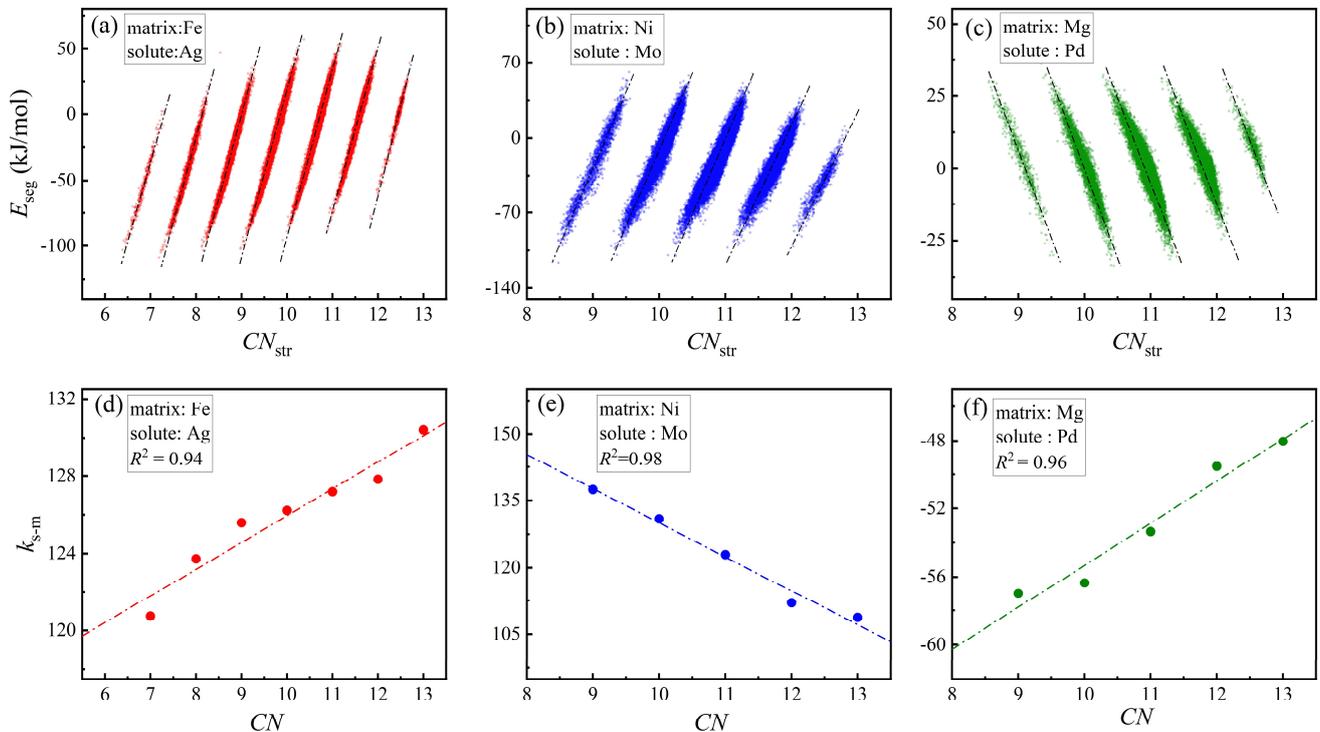

**Figure 1**. The linear functions of $E_{seg}$ and the strained coordination numbers ($CN_{str}$) and their slopes ($k$) in each interval against the usual coordination numbers ($CN$).[14] (a), (d) Ag solute in Fe matrix, (b), (e) Mo solute in Ni matrix, and (c), (f) Pd solute in Mg matrix.

16 different elements Ag, Al, Au, Co, Cu, Fe, Mg, Mo, Ni, Pb, Pd, Pt, Ta, Ti, W and Zr as either solutes or matrices[14]. Compared with the lattice sites of pristine bulk, those at polycrystal GBs experience the change of bond length, bond angle, and coordination number. It is a fundamental challenge to combine these different effects into a simple metric, since the change of bond length and bond angle is associated with strain and that of coordination number with bond breaking/forming.

As the energy of bond stretching is usually much larger than that of bond-angle bending, we mainly focus on the change of bond length towards strain[15,16]. The elastic model was proposed to determine the segregation energies of symmetric GBs[17,18], which is however questionable for the amorphous structures of polycrystal GBs[19]. Indeed, the bond length of 62% bonds at polycrystal GBs of metal matrices ($d_{PGB}$) has been changed by >2% relative to that of bonds at bulk $d_0$, beyond the scope of elastic strain, corresponding to the plastic strain. Namely, the bond length of atoms at polycrystal GBs of the studied metals has been stretched to close to that of alkali metals (AMs). Accordingly, the stretched bond energy ($E_{stb}$) of a given atom at polycrystal GBs of the studied metals likely resembles the behavior of bonds of AMs, which can be represented by a simple Coulombic-like interaction, with $E_{stb,i} \propto 1/d_i$ ($i$ denotes the $i$th first-nearest neighbor atom of a given atom)[20]. With the first-order approximation, one can compute the bonding energy of an atom at polycrystal GBs of the studied metals with its neighboring atoms in a pair-wise way, $E_{bond} = \Sigma E_{stb,i} \propto \Sigma(1/d_i)$.

We combine $d_0$ and $\Sigma(1/d_i)$ to obtain $d_0\Sigma(1/d_i)$, which not only quantifies the strain effects on the bonding energy, but also is another expression of coordination numbers. Namely, we have provided an effective means to unify the concept of plastic strain and bond breaking. We thus propose the strained coordination number $\Sigma(d_0/d_i)$ as a descriptor for determining the segregation energy of solutes at polycrystal GBs,

$$CN_{str} = \sum_{i=1}^{CN} (d_0/d_i) \quad (1)$$

$CN$ is the usual coordination number of GB sites. We find that the segregation energies $E_{seg}$ are linear functions of $CN_{str}$ for all the studied solute-matrix pairs at the specific intervals (see Fig. 1a-c and Supplementary Fig. 1a-c) as,

$$E_{seg} = kCN_{str} + b \quad (2)$$

where $k$ and $b$ are the slope and intercept. The specific intervals can be characterized with $CN$. To further understand $CN_{str}$-$E_{seg}$ linear relations, we adopt the separation variables method, by fixing two of the three variables $CN$, solutes, and matrices and focusing on only one effect of them.

Notably, for a solute-matrix pair, the slope of $CN_{str}$-$E_{seg}$ linear relations, $k_{s-m}$, scales linearly with $CN$ as the intervals change (see Fig. 1d-f and Supplementary Fig. 1d-f), with the regression coefficients above 0.85 as,

$$k_{s-m} = m_{s-m}CN + n_{s-m} \quad (3)$$

where $m_{s-m}$ and $n_{s-m}$ are constants.

We now focus on how the solute effects determine the slope and intercept of the $CN_{str}$-$E_{seg}$ linear relations for a specific coordination-number site in a given matrix ($k_{cs-m}$ and $b_{cs-m}$). Taking Ni, Mg, and Fe matrices at $CN$ =11 as examples, Fig. 2a-c shows that the intercept $b_{cs-m}$ exhibits perfect linear functions against the slope $k_{cs-m}$ with the regression coefficients above 0.99, regardless of the solutes and $CN_{str}$ intervals. In particular, the linear relation between $k_{cs-m}$ and $b_{cs-m}$ is dependent on $CN$. An expression is thus extracted as,

$$b_{cs-m} = -CNk_{cs-m} - 10 \quad (4)$$

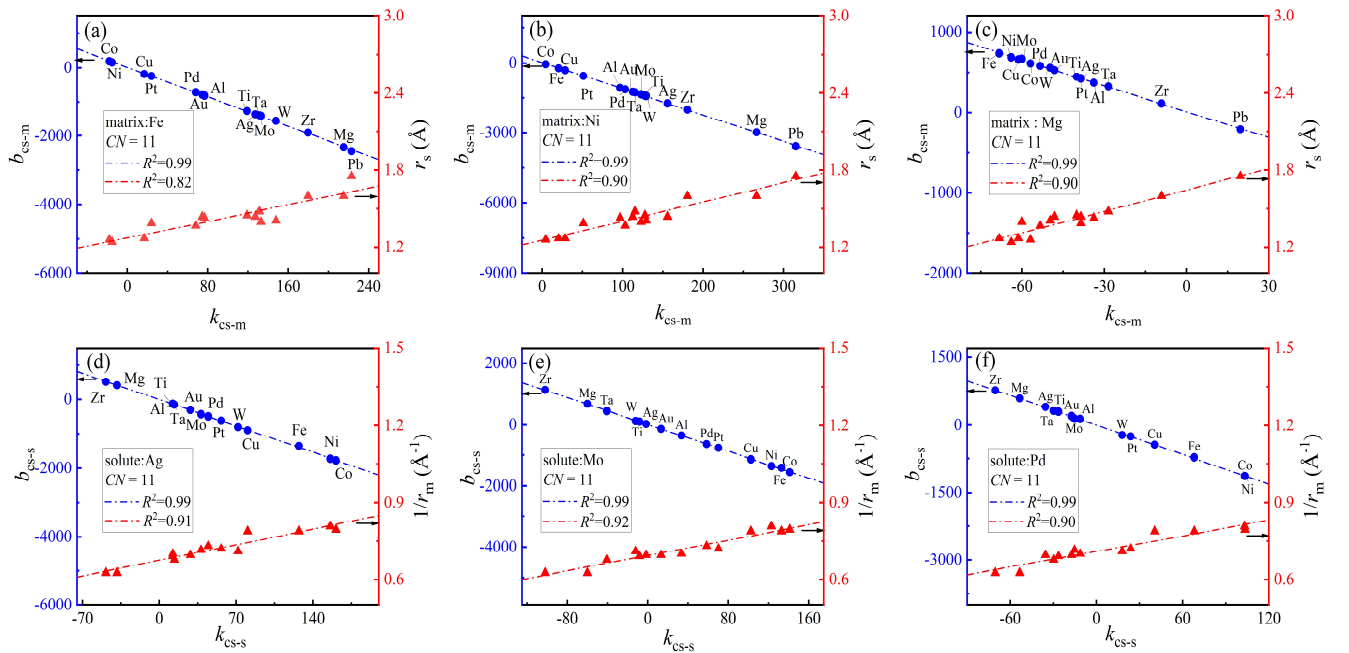

**Figure 2.** Dependence of the slope $k$ and intercept $b$ of $CN_{str}$-$E_{seg}$ functions on the atomic radius of solutes and matrices. (a) Fe, (b) Ni, and (c) Mg matrices with different solutes in the interval of $CN$=11. (d) Ag, (e) Mo, and (f) Pd solutes in different matrices in the interval of $CN$=11. The accuracy is quantified by the regression coefficient $R^2$.

Moreover, the slope $k_{cs-m}$ scales linearly with the atomic radius of solutes ($r_s$), with the regression coefficients above 0.80 (see Fig. 2a-c) as,

$$k_{cs-m} = m_{cs-m}r_s + n_{cs-m} \quad (5)$$

where $m_{cs-m}$ and $n_{cs-m}$ are constants. More results are shown in Supplementary Fig. 2a-c. Therefore, the solute effects of segregation at polycrystal GBs mainly depend on the atomic size of solutes.

Similar with the solute effects, the matrix effects of segregation also strongly depend on the atomic size of matrices. For a given solute, the slope of the $CN_{str}$-$E_{seg}$ linear relations also exhibits linear relationships with the intercept at a specific coordination-number site across various matrices ($k_{cs-s}$ and $b_{cs-s}$), and the slope and intercept are determined by the atomic radius of matrices ($r_m$), with the regression coefficients above 0.80 as illustrated in Fig. 2d-f and Supplementary Fig. 2d-f,

$$k_{cs-s} = m_{cs-s}/r_m + n_{cs-s} \quad (6)$$

where $m_{cs-s}$ and $n_{cs-s}$ are constants.

We now know that the slope of $CN_{str}$-$E_{seg}$ linear relations $k$ linearly depends on $r_s$ and $1/r_m$. By correlating $k_{s-m}$ with the ratio of $r_s$ and $r_m$ ($r_s/r_m$), we again obtain the linear relations, whose slope can be determined by the cohesive energy of matrices ($E_{c,m}$), as $k_{s-m} \propto E_{c,m}(r_s/r_m)$.

Therefore, $r_s$, $r_m$, $CN$, and $E_{c,m}$ together determine the $k$ and $b$ of $CN_{str}$-$E_{seg}$ linear relations. Now, we can estimate the segregation energies of solutes at polycrystal GBs as,

$$E_{seg} = kCN_{str} + b$$
$$= \left[\frac{15}{2}E_{c,m}(30-CN)\left(\frac{r_s}{r_m}-1\right)+10\right](CN_{str} - CN) - 10 \quad (7)$$

$CN_{str}$, $r_s/r_m$, and $CN$ are all dimensionless quantities, and thus $E_{seg}$ has the same unit as $E_{c,m}$. The predicted slope values by the term $\frac{15}{2}E_{c,m}(30-CN)\left(\frac{r_s}{r_m}-1\right)+10$ of Eq. (7) are in good agreement with the calculated ones (see Fig. 3a), demonstrating the robustness of our model.

According to Eq. (7), the term $r_s/r_m-1$ dominates the trend of slope $k$, indicating the importance of the relative size of solutes to matrices in determining the relation between $CN_{str}$ and $E_{seg}$. If $r_s > r_m$ (by 1~5% depending on $CN$), namely for the oversized solutes, the slope of $CN_{str}$-$E_{seg}$, $k$, is positive, otherwise it is negative for the undersized solutes. These findings reflect the fact that the oversized solutes prefer to segregate to the sites with lower $CN_{str}$ in a given interval, since these sites lead to the tensile strain effects and the larger space in comparison with pristine bulk according to Eq. (7). In contrast, the undersized solutes tend to segregate to the sites with larger $CN_{str}$ in a given interval due to the compressive strain effects and smaller sites. These findings are consistent with the previous studies[21-24]. For example, in the symmetric tilt W GBs, the oversized solutes (such as Hf and Ta) prefer to segregate to the stretched sites of GBs, while the undersized elements (such as Os and Ir) tend to segregate to the compressed sites of GBs[21]. Our model thus theorizes

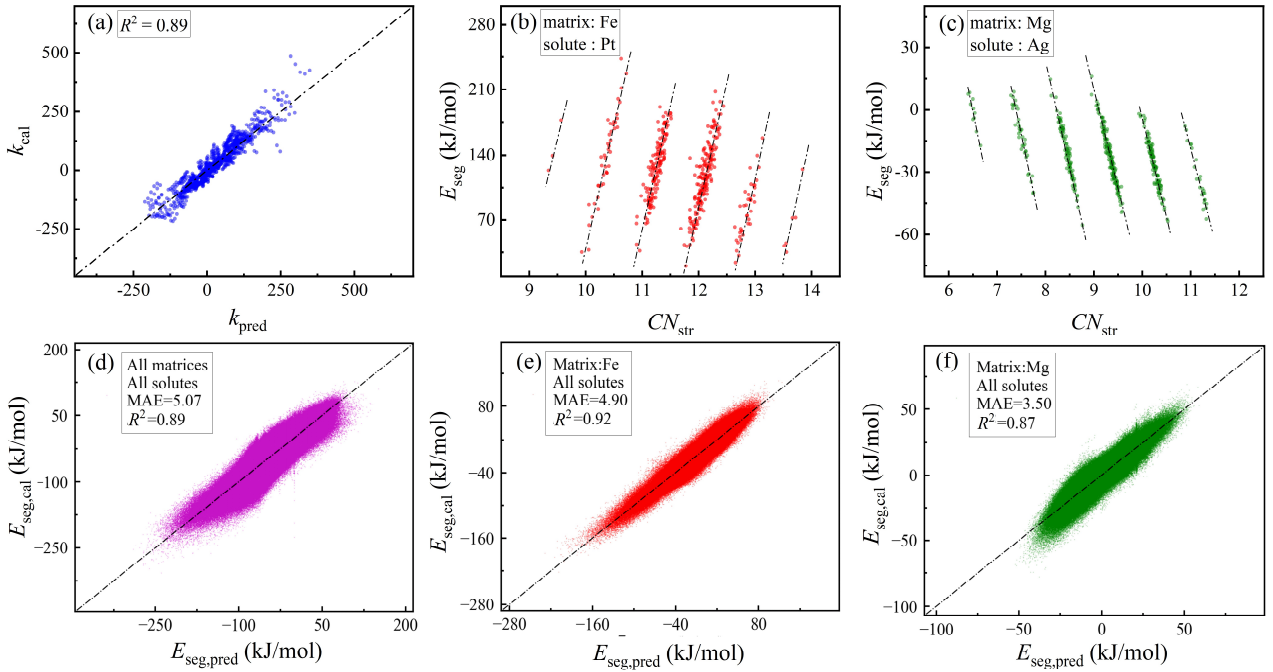

**Figure 3.** The prediction accuracy of our models and their application to bulk metallic glasses (BMGs). (a) The calculated slope $k_{cal}$ of the $CN_{str}$-$E_{seg}$ functions against the predicted slope $k_{pred}$ by Eq. (7). (b), (c) $E_{seg}$ (by DFT calculations) of (b) Pt solute in Fe BMGs and (c) Ag solute in Mg BMGs as a function of $CN_{str}$. (d-f) Comparison of the predicted and calculated values $E_{seg,pred}$ and $E_{seg,cal}$ for (d) all studied matrix-solute pairs, (e) all studied solutes in Fe matrix, and (f) all studied solutes in Mg matrix.

these literature findings regarding the qualitative judgment of separation propensity.

We expect the descriptor $CN_{str}$ to hold in determining the energy of a solute in BMGs, as polycrystal GBs were ever suggested as amorphous states[25]. Our MD, ab intio MD (AIMD), and DFT calculations consistently confirm this expectation, showing that the energies of solutes in BMGs linearly scale with $CN_{str}$ (see Fig. 3b and 3c and Supplementary Fig. 3a). Therefore, our scheme demonstrates the previous assumption that polycrystal GBs are similar with amorphous structures[25], at least in the interatomic bonding, and $CN_{str}$ is general in determining the segregation of solutes in the disordered structures. Furthermore, our scheme reveals that compared with bulks, polycrystal GBs and BGMs exhibit the plastic strain with unusual Coulombic-like bonding behavior, reasonably resembling the trend of the strain-stress curve of TMs in the plastic scope.

Furthermore, the first nearest neighbors of a given site at polycrystal GBs and BMGs are found to be sufficient in determining the segregation energies, indicating that the coupling effect of solutes and matrices at polycrystal GBs and BMGs is highly localized. The screening effect is thus strong in both polycrystal GBs and BMGs. This is different from symmetric tilt GBs, for which the effect of coordination numbers had been found to reach the second-nearest neighbors[26]. Because the lattice sites of symmetric tilt GBs are more consistent with free surfaces (compared with those of polycrystal GBs), where the bond-breaking effects are relatively long-range.

We now study the predictive accuracy of our scheme in determining the segregation energies of solutes at polycrystal GBs and BMGs, by comparing with >2×10$^7$ MD data points and >1000 DFT data points. If one compares all the studied solute-matrix pairs together, the MAEs of our scheme is 5.07 kJ/mol with the segregation-energy spans about 420 kJ/mol (Fig. 3d). For one matrix with all studied solutes, the MAEs of our scheme are below 5 kJ/mol (Fig. 3e and 3f and Supplementary Fig. 3b). If one compares each solute-matrix pair individually, the MAEs of our scheme are generally below 6.5 kJ/mol and often around 1 kJ/mol, namely <5% margin of error (Supplementary Fig. 3c-e). This accuracy is comparable to that of the ML models with high fidelity (Supplementary Table 1) [14]. Notably, these ML models exhibit the complex feature matrix of size with all the GB atoms and many SOAP features ($N^{GB}$ atoms × $F^{SOAP}$ features), impeding the convenient determination of segregation energies and the deep understanding of segregation behavior. Furthermore, these ML models are used to determine segregation energies for a single solute-matrix pair since it only describes the polycrystal-GB effects. In contrast, our framework quantifies not only the polycrystal-GB effects but also the solute and matrix effects.

In summary, we propose an effective descriptor, the strained coordination number, to quantify the role of the complicated structures of polycrystal GBs and amorphous structures in determining the segregation energies of solutes. This descriptor successfully unifies the concept of plastic strain and coordination number, and thus enables us to construct a predictive model for determining the segregation energies of polycrystal GBs across various solutes and matrices, by combining with the usual coordination number, atomic radius of solutes and matrices, and cohesive energy of matrices. This analytic model identifies a linear coupling rule of segregation sites, solutes and matrices for GB segregation in polycrystals, and reveals an unconventional Coulombic-like and localized nature of the solute-matrix bonding at polycrystal GBs and BMGs. Our framework thus provides a deep physical insight into GB segregation in polycrystals and a predictive tool of high-performance-alloy design.

**References**
1. Wood, D. L. & Westbrook, J. H. Embrittlement of Grain Boundaries by Equilibrium Segregation. *Nature* **192**, 1280-1281 (1961).
2. Duscher, G., Chisholm, M. F., Alber, U. & Rühle, M. Bismuth-induced embrittlement of copper grain boundaries. *Nat. Mater.* **3**, 621-626 (2004).
3. King, A. et al. Observations of intergranular stress corrosion cracking in a grain-mapped polycrystal. *Science* **321**, 382-385 (2008).
4. Mishin, Y., Asta, M. & Li, J. Atomistic modeling of interfaces and their impact on microstructure and properties. *Acta Mater.* **58**, 1117-1151 (2010).
5. Lejcek, P., Sob, M. & Paidar, V. Interfacial segregation and grain boundary embrittlement: An overview and critical assessment of experimental data and calculated results. *Prog. Mater. Sci.* **87**, 83-139 (2017).
6. McLean, D. *Grain boundaries in metals*. (Oxford University Press, Oxford, 1957).
7. Lejcek, P. *Grain boundary segregation in metals*. (Springer Science & Business Media, 2010).
8. Tran, R. et al. Computational study of metallic dopant segregation and embrittlement at molybdenum grain boundaries. *Acta Mater.* **117**, 91-99 (2016).
9. Razumovskiy, V. I., Lozovoi, A. Y. & Razumovskii, I. M. First-principles-aided design of a new Ni-base superalloy: Influence of transition metal alloying elements on grain boundary and bulk cohesion. *Acta Mater.* **82**, 369-377 (2015).
10. Mahjoub, R., Laws, K. J., Stanford, N. & Ferry, M. General trends between solute segregation tendency and grain boundary character in aluminum - An ab initio study. *Acta Mater.* **158**, 257-268 (2018).
11. Wagih, M. & Schuh, C. A. Spectrum of grain boundary segregation energies in a polycrystal. *Acta Mater.* **181**, 228-237 (2019).
12. Wagih, M. & Schuh, C. A. The spectrum of interstitial solute energies in polycrystals. *Scripta Mater.* **235**, 115631 (2023).
13. Wagih, M. & Schuh, C. A. Learning Grain-Boundary Segregation: From First Principles to Polycrystals. *Phys. Rev. Lett.* **129** (2022).
14. Wagih, M., Larsen, P. M. & Schuh, C. A. Learning grain boundary segregation energy spectra in polycrystals. *Nat. Commun.* **11**, 6376 (2020).
15. Sun, C. Q. et al. Length, strength, extensibility, and

thermal stability of a Au-Au bond in the gold monatomic chain. *J. Phys. Chem. B* **108**, 2162-2167 (2004).
16. Persaud, R. R., Chen, M. Y., Peterson, K. A. & Dixon, D. A. Potential Energy Surface of Group 11 Trimers (Cu, Ag, Au): Bond Angle Isomerism in Au. *J. Phys. Chem. A* **123**, 1198-1207 (2019).
17. Huber, L., Rottler, J. & Militzer, M. Atomistic simulations of the interaction of alloying elements with grain boundaries in Mg. *Acta Mater.* **80**, 194-204 (2014).
18. Huber, L. et al. Ab initio modelling of solute segregation energies to a general grain boundary. *Acta Mater.* **132**, 138-148 (2017).
19. Li, J. et al. Atomistic mechanisms governing elastic limit and incipient plasticity in crystals. *Nature* **418**, 307-310 (2002).
20. Courtney, T. H. *Mechanical Behavior of Materials*. 2nd ed. edn (Waveland Press, Inc., Long Grove, IL, 2005).
21. Wu, X. B. et al. First-principles determination of grain boundary strengthening in tungsten: Dependence on grain boundary structure and metallic radius of solute. *Acta Mater.* **120**, 315-326 (2016).
22. Huang, Z. F. et al. Understanding solute effect on grain boundary strength based on atomic size and electronic interaction. *Sci. Rep.* **10**, 16856 (2020).
23. Lei, C., Xue, H. T., Tang, F. L. & Luo, X. The effect of solute segregation on stability and strength of Cu symmetrical tilt grain boundaries from the first-principles study. *J. Mater. Res. Technol.* **21**, 3274-3284 (2022).
24. Ito, K. & Sawada, H. First-principles analysis of the grain boundary segregation of transition metal alloying elements in γFe. *Comput. Mater. Sci.* **210**, 111050 (2022).
25. Chandross, M. & Argibay, N. Ultimate Strength of Metals. *Phys. Rev. Lett.* **124**, 125501 (2020).
26. Li, X., Li, Y. & Gao, W. An analytic descriptor for determining the effect of grain-boundary structures of metals on solute segregation. *J. Appl. Phys.* **135**, 145303 (2024).

## Method

### Polycrystal structures

In this study, we analyze the polycrystal structures and segregation energy data from Ref.[14]. The calculations are performed using the atomistic simulation package LAMMPS[27] for Molecular statics (MS) and molecular dynamics (MD) simulations. Polycrystal structures are built with Atomsk[28] by filling a 20*20*20 nm$^3$ volume with 16 randomly oriented grains and ~10$^5$ segregation sites for each matrix. Under a Nose-Hoover thermostat/barometric for 250 ps using a time step of 1 fs, grain structure and boundaries were thermally annealed at 0.3~0.5 of melting point to relax without permitting exaggerated grain growth. It is then slowly cooled to 0K at a cooling rate of 3K/ps, and finally reaches the conjugate gradient energy minimization. These polycrystal GBs contain low-angle boundaries or subpopulations of the polycrystalline GB microenvironments such as triple junctions and quadruple nodes. The adopted polycrystal GB data thus exhibit the complexity of solute and matrix effects as well as the diverse geometric environments of segregation sites, which have been proven to be effective in determining the segregation spectrum of polycrystals. We refer the readers to Ref. [14] for more details of the polycrystal structures.

### The segregation energies of polycrystal GBs

Segregation energies ($E_{seg}$) quantitatively describe the driving force for the segregation of a solute atom at GBs, which are calculated by the following formula:

$$E_{seg} = E_{GB}^{solute} - E_{bulk}^{solute} \tag{1}$$

where $E_{GB}^{solute}$ is the internal energy of GB structures including solute atoms, and $E_{bulk}^{solute}$ is the internal energy of bulk structures including solute atoms.

### Bulk metallic glasses (BMGs) with density functional theory calculations

For the ab initio MD (AIMD) simulations, we use canonical NPT ensembles under the Langevin thermostat with the Vienna ab initio simulation package (VASP)[29]. All calculations are performed using Projected augmented planewaves (PAW) with the Perdew-Burke-Ernzerhof exchange-correlation potentials[30]. The 350 eV cutoff energies and gamma point are chosen. To obtain amorphous metal structures, the metal bulk structures about 400-500 atoms are melted and equilibrated at the temperature of 100-200 K than the melting point temperatures for 2500 timesteps, with each timestep of 4 fs. We cool the systems to 0 K over a period of 10 ps, using a timestep of 4 fs, to achieve amorphous structures. Subsequently, we replace every atom in the structure with a solute atom and calculate the single-point energy. The segregation energy of solutes at BMGs is calculated by the formula as,

$$E_{seg} = (E_{amo}^{solute} - E_{amo}) - (E_{bulk}^{solute} - E_{bulk}) \tag{2}$$

where $E_{amo}^{solute}$ is internal energy of amorphous structure including solute atom, $E_{amo}$ is internal energy of amorphous structure without solute atom, $E_{bulk}^{solute}$ is internal energy of bulk structure including solute atom, and $E_{bulk}$ is internal energy of bulk structure without solute atom. Matrix metals selected include Fe, Ag, and Mg.

### Strained coordination numbers

As polycrystal GBs experience serious lattice distortion and are even in amorphous states as BMGs, it is critical and necessary to set a cutoff radius in order to count the first nearest neighbors of central sites. Within the cutoff radius of a given central atom, all neighbors are designated as the nearest ones. Notably, the cutoff radius of the nearest neighbors impacts the calculations of $CN_{str}$ but weakly affects the linear relationship of $E_{seg}$ and $CN_{str}$. $CN_{str}$ always linearly scales with the segregation energies for various solute-matrix pairs at the cutoff radius of $1.10d_0$~$1.20d_0$, with the optimal accuracy at $1.15d_0$ as shown in Supplementary Fig. 4. Therefore, a cutoff radius of $1.15d_0$ is adopted based on the extensive tests. All these results show that $CN_{str}$ is effective in determining the segregation energies across various polycrystal-GB sites, regardless of the type of solutes and matrices. We compile $CN_{str}$ into a Python code in order to rapidly and conveniently apply to the polycrystal structures with almost ~10$^5$ atoms.

### References

27. Thompson, A. P. et al. LAMMPS-a flexible simulation tool for particle-based materials modeling at the atomic, meso, and continuum scales. *Comput. Phys. Commun.* **271**, 108171 (2022).
28. Hirel, P. Atomsk: A tool for manipulating and converting atomic data files. *Comput. Phys. Commun.* **197**, 212-219 (2015).
29. Kresse, G. & Furthmüller, J. Efficient iterative schemes for ab initio total-energy calculations using a plane-wave basis set. *Phys. Rev. B* **54**, 11169-17979 (1996).
30. Perdew, J. P., Burke, K. & Ernzerhof, M. Generalized gradient approximation made simple,. *Phys. Rev. Lett.* **77**, 3865-3868 (1996).

### Competing interests

The authors declare no competing interests.

### Additional information

Supplementary information is available for this paper at https://doi.org/xxx.

### Acknowledgments

The authors are thankful for the support from the National Natural Science Foundation of China (Nos. 22173034, 11974128, 52130101), the Opening Project of State Key Laboratory of High Performance Ceramics and Superfine

**Author information**

*Email: wgao@jlu.edu.cn

*Email: jiangq@jlu.edu.cn

**Author contributions**

W.G. and Q.J. conceived the original idea and designed the strategy. X.L. collected the data and H.W. performed the calculations. W.G. and X.L. derived the models and analyzed the results with the contribution from H.W. W.G., X.L., and H.W. together wrote the manuscript. H.W. prepared the Supplementary Information and drew all figures with the contribution from X.L. All authors have discussed and approved the results and conclusions of this article.

**Competing interests**

The authors declare no competing interests.